\documentclass[sigconf,natbib]{acmart}
\AtBeginDocument{%
  \providecommand\BibTeX{{%
    \normalfont B\kern-0.5em{\scshape i\kern-0.25em b}\kern-0.8em\TeX}}}

\setcopyright{acmcopyright}
\copyrightyear{2018}
\acmYear{2018}
\acmDOI{XXXXXXX.XXXXXXX}

\acmConference[Conference acronym 'XX]{Make sure to enter the correct
  conference title from your rights confirmation emai}{June 03--05,
  2018}{Woodstock, NY}
\acmPrice{15.00}
\acmISBN{978-1-4503-XXXX-X/18/06}

\begin{document}

\title{GPT4Rec: A Generative Framework for Personalized Recommendation and User Interests Interpretation}

\author{Jinming Li$^1$, Wentao Zhang$^2$, Tian Wang$^2$, Guanglei Xiong$^2$, Alan Lu$^2$, Gerard Medioni$^2$}
\email{lijinmin@umich.edu, {wentazha,wangtan,glx,alalu,medioni}@amazon.com}
\affiliation{\institution{ $^1$University of Michigan, Ann Arbor \quad\quad\  $^2$Amazon, United States\country{ }}}


% \affiliation{%

% \affiliation{%

% \affiliation{%

\renewcommand{\shortauthors}{Li, et al.}

%--------------------------------------------------------------------------------------------------------
%% The abstract is a short summary of the work to be presented in the article.
\begin{abstract}

Recent advancements in Natural Language Processing (NLP) have led to the development of NLP-based recommender systems that have shown superior performance. However, current models commonly treat items as mere IDs and adopt discriminative modeling, resulting in limitations of (1) fully leveraging the content information of items and the language modeling capabilities of NLP models; (2) interpreting user interests to improve relevance and diversity; and (3) adapting practical circumstances such as growing item inventories. To address these limitations, we present GPT4Rec, a novel and flexible generative framework inspired by search engines. It first generates hypothetical "search queries" given item titles in a user's history, and then retrieves items for recommendation by searching these queries. The framework overcomes previous limitations by learning both user and item embeddings in the language space. To well-capture user interests with different aspects and granularity for improving relevance and diversity, we propose a multi-query generation technique with beam search. The generated queries naturally serve as interpretable representations of user interests and can be searched to recommend cold-start items. With GPT-2 language model and BM25 search engine, our framework outperforms state-of-the-art methods by $75.7\%$ and $22.2\%$ in Recall@K on two public datasets. Experiments further revealed that multi-query generation with beam search improves both the diversity of retrieved items and the coverage of a user's multi-interests. The adaptiveness and interpretability of generated queries are discussed with qualitative case studies.

\end{abstract}

\begin{CCSXML}
<ccs2012>
<concept>
<concept_id>10002951.10003317.10003347.10003350</concept_id>
<concept_desc>Information systems~Recommender systems</concept_desc>
<concept_significance>500</concept_significance>
</concept>
<concept>
<concept_id>10002951.10003317.10003338.10003341</concept_id>
<concept_desc>Information systems~Language models</concept_desc>
<concept_significance>500</concept_significance>
</concept>
</ccs2012>
\end{CCSXML}

\ccsdesc[500]{Information systems~Recommender systems}
% \ccsdesc[500]{Information systems~Language models}

\keywords{personalized recommendation, user interests modeling, generative language models, query generation, searching}

\maketitle

%--------------------------------------------------------------------------------------------------------
%--------------------------------------------------------------------------------------------------------

\section{Introduction}

Recommender systems are information filtering systems that provide personalized suggestions for items that are relevant to a particular user, ubiquitously adopted in applications including E-commerce and social media services \cite{rendle2010factorization, covington2016deep, DBLP:conf/recsys/Kula15, pal2020pinnersage}. With the development of Natural Language Processing (NLP), many NLP-based models are proposed for personalized recommendation by modeling the user-item interaction sequences \cite{jannach2017recurrent, wu2018starspace, zhou2020s3}. The recent success of Large Language Models (LLM, \cite{vaswani2017attention, devlin2018bert, radford2019language, brown2020language}) that achieved superior performances across various NLP tasks or diagolue (ChatGPT \cite{radford2018improving}), have spurred the research of LLM-based recommender systems\cite{sun2019bert4rec, geng2022recommendation, kang2018self, sileo2022zero, zhang2021language, penha2020does}. In particular, BERT4Rec \cite{sun2019bert4rec} adopted the bi-directional self-attention structure of BERT \cite{devlin2018bert} and outperformed other NLP-based models and sequential models\cite{ranzato2015sequence, jannach2017recurrent, kang2018self}.

Despite improvement in model performances, existing NLP-based models \cite{sun2019bert4rec, geng2022recommendation, kang2018self, sileo2022zero, zhang2021language, penha2020does} typically treat items as mere IDs and adopt discriminative modeling, which have the following limitations: First, these models are incapable of fully leveraging the content information of items and the language modeling ability of NLP models. Second, these models are unable to accommodate changing and growing item inventories, both of which are important practical issues \cite{lam2008addressing, schein2002methods}. Moreover, discriminative models are challenging to interpret user interests, which is essential to improve diversity and quality of recommendation \cite{zhang2020explainable,cen2020controllable,li2019multi,chen2021values}.

To address these limitations, we propose GPT4Rec, a novel and flexible generative framework for personalized recommendation that simultaneously offers interpretations of user interests. Our framework is inspired by search engines: GPT4Rec first generates hypothetical "search queries" with a generative language model, which takes the item titles in a user's history combined with a generation prompt as model input. Then, GPT4Rec retrieves items for recommendation by searching the generated queries with a search engine. The key component of GPT4Rec is a powerful generative language model that learns both user and item embeddings in the language space, allowing us to utilize the semantic information in item titles and capture user's diverse interests. To decode user's multi-interests and improve recommendation diversity, we adopt the multi-query beam search technique in query generation. These queries are human-understandable and hold standalone value for interpreting user interests. In addition, searching queries that represent user interests for recommendation can naturally solve the item cold-start problem and changing item inventory issue. Last but not least, GPT4Rec is favorable in practice as it is flexible to accommodate more advanced LLMs or search engines to improve performance.

In summary, our contributions include: (1) We propose GPT4Rec, a novel and flexible generative framework that regard recommendation task as query generation + searching; (2) We adopt the multi-query beam search strategy to produce diverse and interpretable user interests representations; (3) We conduct experiments on two public datasets, demonstrating substantial improvement of Recall@$K$ against state-of-the-art methods; (4) With both quantitative and qualitative analysis, we investigate and show that the number of generated queries improves the diversity of retrieved items as well as the coverage of users' multi-interests.

\section{Methodology}

%--------------------------------------------------------------------------------------------------------

\begin{figure}[]
    \centering
    \includegraphics[width=\linewidth]{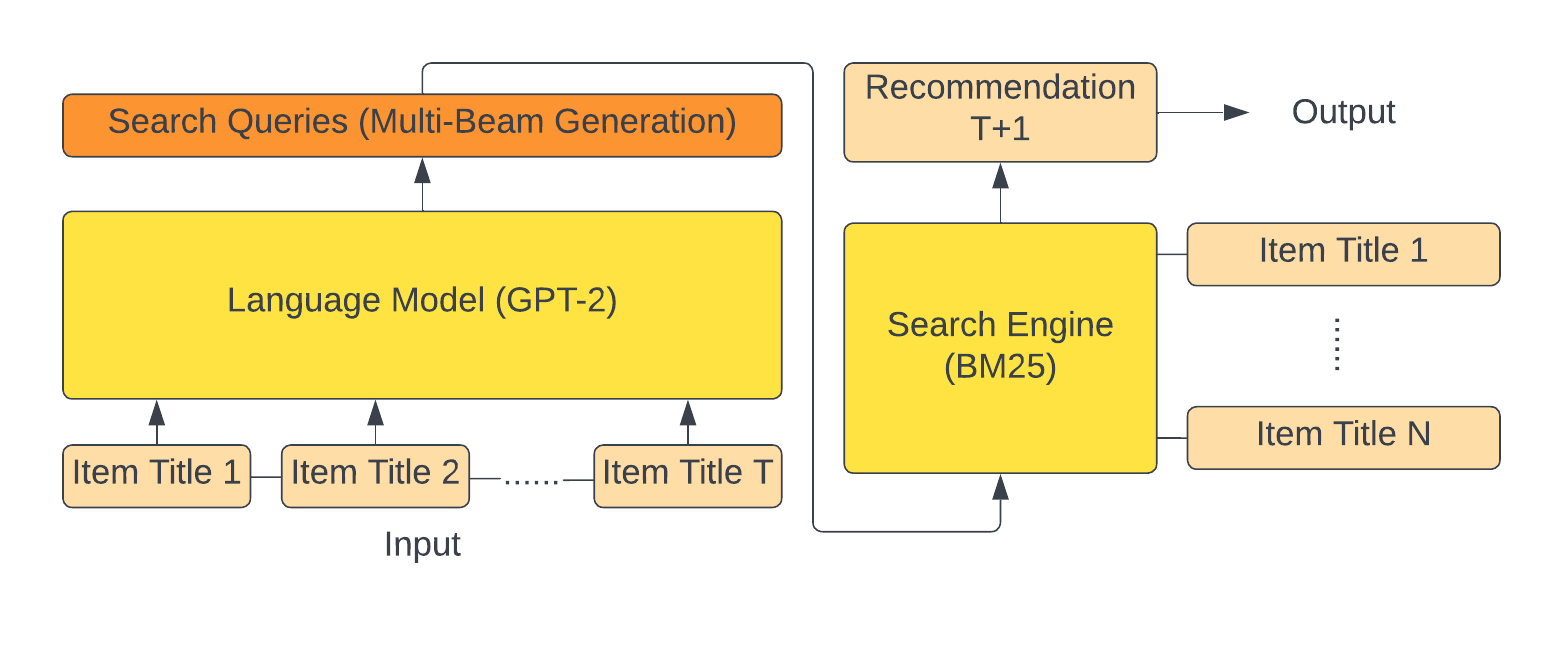}
    \caption{The architecture of the GPT4Rec framework.}
    \label{fig:flow_chart}
\end{figure}

The architecture of the proposed framework is illustrated in Figure \ref{fig:flow_chart}. Given a user's item interaction sequence, GPT4Rec formats the item titles with a prompt and uses a generative language model to learn both item and user embeddings in the language space. The model then generates multiple queries that represent user's interests, which will be fed to a search engine to retrieve items for recommendation. In this paper we choose the GPT-2 language model and BM25 search engine, while the framework is flexible to incorporate more advanced models.

% \subsection{Problem Setup}

% \label{sec:problem_setup}

% We start by formally introducing the notations and the candidate matching problem setup. We use $\mathcal{I}$ to denote the item inventory containing all the the item indexes and $i \in \mathcal{I}$ to denote a specific item index. Similarly, $\mathcal{U}$ denotes the set of all user indexes and $u \in \mathcal{U}$ denotes a specific user index. Each user $u \in \mathcal{U}$ corresponds to a sequence of items that were previously purchased or interacted, denoted as $H^u = \left(h^u_1, h^u_2, ... , h^u_{T_u}\right) \in \mathcal{I}^{T_u}$ with the subscripts $1, 2, ... , T_u$ as the timestamps. We consider the \textit{non-repeatable recommendation} setting, where each item in $H_u$ only appears once and the target item will not appear in user history. Each item $i \in \mathcal{I}$ corresponds to a title, $S(i)$, which consists of a series of words.

% The candidate matching problem under the non-repeatable setting can be formulated as follows. For each user $u \in \mathcal{U}$, given $H^u$, we aim to find $K$ most relevant items from the rest of items in the inventory $\mathcal{I} \setminus H^u$ as the recommendation set for next timestamp $T_u + 1$. We denote $K$ as the number of retrieved items and $R^u = \left(r^u_1, r^u_2, ... , r^u_{K}\right) \in \left(\mathcal{I} \setminus H^u \right) ^K$ as the retrieved item set.

%--------------------------------------------------------------------------------------------------------
\subsection{Query Generation with the Language Model}
\label{sec:query_generation}

The first component of GPT4Rec is a generative language model, whose goal is to learn the user representation in the language space from the item interaction sequence, and then generates multiple queries that represent user interests. Fine-tuning the chosen GPT-2 model, which has 117M parameters with sophisticated transformer-type architecture and is pre-trained on enormous language corpus, serves our purpose of capturing user interests and item content information. Through experiments we decide to use the following prompt to format model input:
\begin{center}
    \textit{Previously, the customer has bought:}
    \newline
    \textit{<ITEM TITLE 1>. <ITEM TITLE 2>... }
    \newline
    \textit{\noindent In the future, the customer wants to buy }
\end{center}
which contains semantic information in item titles and is denoted as $W^u$ for each user $u$. GPT-2 learns user representation from $W_u$ and is then able to generate queries based on the conditional distribution $P(\cdot|W_u)$ in a sequential way. 

In order to well-represent users' diverse interests and increase the diversity of recommendation results, we propose to generate not single but multiple queries with the beam search technique \cite{ranzato2015sequence, vijayakumar2016diverse}. Given beam size $m$, generation score function $S(\cdot)$, and the candidate queries $Q^{l} = (q^{l}_1, q^{l}_2, ... q^{l}_m)$ with length $l$, the beam search algorithm updates $Q^{l+1}$ to be the top-$m$ queries of length $l+1$ that maximize 
$\text{S}(W^u, q), \quad q \in \{ |q| = l+1, q_{[1:l]} \in Q^{l}\}.$
Most importantly, such beam search strategy generates user interests representation that are diverse in different aspects and granularity. 

\subsection{Item Retrieval with the Search Engine}

The second component of the proposed framework is a search engine that functions as a "discriminator". It takes each generated query as input and retrieves the most relevant items in the inventory as output with matching score functions. The matching score functions measure the similarity on the language space, playing a similar role of inner-product similarity in vector-embedding methods \cite{rendle2010factorization, covington2016deep}. We choose to adopt BM25 matching score function \cite{robertson2009probabilistic}, as it is one of the most widely used baseline search engines that accounts for the term frequency saturation and the document length with two corresponding parameters $k_1$ and $b$.

Let $K$ represent the total number of recommendation items and $m$ represent the number of generated queries, we propose a ranking-based strategy to combine searching results of each query. It first retrieves top-$K/m$ items from searching results with the query having highest generation score, then sequentially adds $K/m$ non-repeated items from the rest of queries by their score rankings. The proposed strategy is able to balance both relevance and diversity of the retrieved items.

\subsection{Training Strategy}

We propose a flexible two-step training procedure that optimizes the language model and the search engine separately, which is favorable in practice for model iteration. Given the item interaction sequence $i_1, i_2, ..., i_T$ of each user $u$, we take the first $T-1$ item titles in the prompt shown in Section \ref{sec:query_generation}, then concatenate it with the title of last item $i_T$ to form training corpus to fine-tune the pre-trained GPT-2 model. This strategy is motivated by the contrastive learning literature \cite{jaiswal2020survey}, with the underlying idea that \textit{the most fine-grained and accurate search query is the target item title itself}. After the language model is trained, we optimize the BM25 parameters by grid-searching parameters $k_1$ and $b$ that gives the best performance when searching the generated queries to retrieve target items.

\section{Experiments}

%--------------------------------------------------------------------------------------------------------
\subsection{Experiment Setup}

\subsubsection{Data}

We conduct experiments on two public datasets, 5-core Amazon Review data\cite{he2016ups} in categories Beauty and Electronics. Items with missing or noisy titles (more than $400$ characters) are discarded. Item interaction sequence of each user is deduplicated and truncated at the maximum length of 15. Descriptive statistics of the preprocessed datasets are presented in Table \ref{tab:data_statisitcs}. Following the next-item prediction task setup \cite{sun2019bert4rec}, users sequences are split into training, validation, and test sets by 0.8/0.1/0.1 proportions and the last item in each test sequence is treated as prediction target.

% Each dataset is split into training, validation, and test sets by splitting the user sequences into 0.8/0.1/0.1 proportions. For each user sequence in the training set, a training sample is constructed by treating the latest item as target and the rest of previous items as user history. For each user sequence in the test set, the latest item is treated as the target and the rest of user sequence is used for constructing a training sample.
\begin{table*}[]
  \caption{Overall performance of baseline methods and the proposed framework with different number of generated queries. The best performance is highlighted in bold font and the best baseline results is underlined.} 
  \label{tab:recalls}
  \small
  \begin{tabular}{llcccccccc}
  % \begin{tabular}{llllllllll}
  \toprule
    Dataset     & Recall@$K$    & FM-BPR & ContentRec & YouTubeDNN & BERT4Rec & \multicolumn{4}{c}{GPT4Rec}                                             \\
                &           &         &           &            &          & 5 Queries        & 10 Queries       & 20 Queries       & 40 Queries       \\
    \midrule
    Beauty      & K=5  & 0.0356 & 0.0254    & \underline{0.0376}    & 0.0355  & \textbf{0.0653} &          ---        &       ---           &      ---            \\
                & K=10 & 0.0499 & 0.0440   & \underline{0.0549}    & 0.0513  & 0.0810          & \textbf{0.1036} &          ---        &         ---         \\
                & K=20 & 0.0716 & 0.0644    & 0.0753    & \underline{0.0816}  & 0.1027          & 0.1252          & \textbf{0.1454} &        ---          \\
                & K=40 & 0.1040 & 0.0952   & 0.1066    & \underline{0.1161}  & 0.1297          & 0.1522          & 0.1743          & \textbf{0.2040} \\
    \midrule
    Electronics & K=5  & 0.0345 & 0.0241   & 0.0352     & \underline{0.0362}  & \textbf{0.0434} &        ---          &          ---        &          ---        \\
                & K=10 & 0.0387 & 0.0307   & 0.0435    & \underline{0.0451}  & 0.0480          & \textbf{0.0545} &           ---       &         ---         \\
                & K=20 & 0.0441 & 0.0391   & 0.0539     & \underline{0.0573}  & 0.0524          & 0.0607          & \textbf{0.0705} &        ---          \\
                & K=40 & 0.0505 & 0.0494   & 0.0684     & \underline{0.0751}   & 0.0574          & 0.0672          & 0.0788          & \textbf{0.0918} \\
    \bottomrule
\end{tabular}
\end{table*}

\begin{table}[]
  \caption{Dataset statistics.}
  \label{tab:data_statisitcs}
  \footnotesize
  \begin{tabular}{cccccc}
    \toprule
    Name & \#User & \#Item & \#Interaction & Ave. Length & Meta Info. \\
    \midrule
    Beauty & 22,254 & 11,778 & 190,726 &  7.439 & Cate.\\
    Electronics & 728,719 & 159,456 & 6,724,382 & 7.797 & Cate., Brand\\
    % Clothing & 1,219,677 & 373,539 & 11,285,464 & 8.133\\
    \bottomrule
  \end{tabular}
\end{table}

\subsubsection{Evaluation Metrics} \textbf{Recall@$K$} is the primary metric of interests for next-item prediction, which measures whether the top-$K$ recommended items contains the target item, averaged over all users. Besides Recall@$K$, we are also interested in: 

\noindent\textbf{Diversity@$K$} \cite{chen2021values}, which measures dissimilarity if items as
\begin{center}
    $\text{Diversity@}K = \text{Average}\left[1 - \frac{\sum_{i_1 \neq i_2} \text{Sim}(i_1, i_2)}{K(K-1)}\right],$
\end{center}
    where we adopt the \textit{Jaccard similarity} \footnote{https://en.wikipedia.org/wiki/Jaccard\_index} on item category or brand.

\noindent\textbf{Coverage@$K$}, which measures how well the recommended items $R$ cover the user sequence $U$ in terms of category or brand:
\begin{center}
    $\text{Coverage@}K = \text{Average}\left[\frac{\left|\left(\bigcup_{i \in R} Cate_i\right) \cap \left(\bigcup_{i \in U} Cate_i\right)\right|}{\left|\bigcup_{i \in U} Cate_i\right|}\right],$
\end{center}
    which is a more reasonable metric for measuring the coverage of user's multiple interests as Diversity@$K$ would favor totally random/irrelevant recommendation.
    
    % Under Diversity@$K$, which has been used in existing literature of multi-interest recommender systems \cite{li2019multi,cen2020controllable}, a totally random recommendation set can still achieve high diversity score even if it has zero relevance with user's interests.
% $$\text{Recall@}K = \text{Ave} \left[ \mathbb{I}\left(H^u_{T_{u}} \in R^u\right)\right],$$
% where the average is taken over all users in the test set, $\mathbb{I}(\cdot)$ is the indicator function, with $H^u_{T_u}$ as the target item and $R^u$ as the retrieved item set.

\subsubsection{Baseline Methods} 

\noindent\textbf{FM-BPR} \cite{DBLP:conf/recsys/Kula15} is a Factorization Machine model with Bayesian Personalized Ranking criteria, which learns the collaborative information from user-item interaction matrix. 

\noindent\textbf{ContentRec} is a content-based model widely used in industry, which learns bad-of-words embeddings from item titles and mean-pooling over user-interacted items for user embeddings. 

\noindent\textbf{YouTubeDNN} \cite{covington2016deep} is one of the most popular deep-learning model used in industry, which adopts a neural network to learn user/item embeddings.

\noindent\textbf{BERT4Rec} \cite{sun2019bert4rec} is a state-of-the-art BERT-based model, with bidirectional self-attention architecture and masked training strategy.

\subsubsection{Implementation Details}
We implement the GPT-2 model released by HuggingFace \cite{wolf2019huggingface} with 117M parameters and fine-tune the model for 20 epochs using the Adam optimizer with weight decay. We set the learning rate to be $0.0001$ and initialize the optimizer with 2000 warm-up steps. BM25 is optimized over $k1 \in [0,3], \quad b \in (0,1)$ with previously discussed method. ContentRec is implemented using StarSpace \cite{wu2018starspace} to learn item title embeddings and the other baseline methods are implemented with public available Github repositories \footnote{LightFM: https://github.com/lyst/lightfm \\ StarSpace: https://github.com/facebookresearch/StarSpace \\ DeepMatch: https://github.com/shenweichen/DeepMatch \\ RecBole: https://github.com/RUCAIBox/RecBole}. For all the baseline methods, we select the optimal embedding dimensions from $64, 128, 256$ on the validation set while setting the rest of hyperparameters following the authors' suggestions in their original papers.

%--------------------------------------------------------------------------------------------------------
\subsection{Quantitative Analysis}

\subsubsection{Overall Performance} Table \ref{tab:recalls} summarizes the Recall@$K$ of GPT4Rec with varying numbers of generated queries compared to baseline methods on two datasets. Our proposed framework outperforms all baseline methods on both datasets, achieving a $\textbf{75.7\%}$ relative improvement on the smaller-scaled Beauty dataset in Recall@40, and $\textbf{22.2\%}$ on the larger-scaled Electronics dataset. 

The comparison with baseline methods suggests that both item content information and modern language modeling are key ingredients for achieving superior performance. One the one hand, while BERT4Rec has the best performance among the baseline methods by leveraging modern language modeling techniques, it fails to fully utilize the item content information by treating items as IDs. On the other hand, ContentRec's use of item content information with bag-of-words embeddings and mean-pooling modeling is insufficient for achieving comparable performance.

\subsubsection{Advantages of Multi-query Generation}

When looking at the lower-triangle numbers of both datasets shown in Table \ref{tab:recalls}, monotone increasing trends can be found in row-wise, column-wise, and diagonal-wise directions. These trends indicate that the multi-query beam search generation strategy greatly improves the relevance of recommended items. In particular, generating $K$ queries and retrieving one item per query yields the best performance of Recall@$K$. This finding suggests that \textit{each query contains enough detail to retrieve a relevant item}.

We further investigate the diversity and coverage of user interests in recommended items, as well as their relationship to the multi-query generation strategy. As reported in Table \ref{tab:diveristy}, for both Diversity@$K$ and Coverage@$K$ the best performance are achieved by generating $K$ queries and retrieve one item per query, which aligns our finding for Recall@$K$ in Table \ref{tab:recalls}. Similar monotone increasing trends indicate that multi-query generation strategy produces more comprehensive user interests representations. On the other hand, as we see in columns of Diversity@$K$, increasing $K$ without increasing the number of queries will not improve diversity. This shows searching doesn't assist in improving diversity.

\begin{table*}[]
    \caption{Diversity and coverage of user interests versus the number of generated queries. Highest values with regard to category/brand information are highlighted in bold font for two datasets.} 
    \label{tab:diveristy}
    \small
    \begin{tabular}{cccccccccccccc}
    \toprule
    Dataset           &      & \multicolumn{4}{c}{Beauty (Category)}                             & \multicolumn{4}{c}{Electronics (Category)}                        & \multicolumn{4}{c}{Electronics (Brand)}                           \\
    \midrule
    Number of Queries &      & 5              & 10             & 20             & 40             & 5              & 10             & 20             & 40             & 5              & 10             & 20             & 40             \\
    \midrule
    Diversity@K       & K=5  & \textbf{0.679} & ---      &         ---       &           ---     & \textbf{0.671} &         ---       &     ---           &      ---          & \textbf{0.534} &         ---       &       ---         &    ---            \\
    \textbf{}         & K=10 & 0.654          & \textbf{0.716} & ---      &         ---       & 0.617          & \textbf{0.733} &        ---        &         ---       & 0.529          & \textbf{0.643} &      ---          &        ---        \\
                      & K=20 & 0.659          & 0.706          & \textbf{0.749} & ---     & 0.605          & 0.703          & \textbf{0.778} &        ---        & 0.559          & 0.645          & \textbf{0.717} &          ---      \\
                      & K=40 & 0.679          & 0.715          & 0.749          & \textbf{0.783} & 0.601          & 0.696          & 0.762          & \textbf{0.811} & 0.604          & 0.669          & 0.724          & \textbf{0.767} \\
    \midrule
    Coverage@K        & K=5  & \textbf{0.417} & ---      &          ---      &        ---        & \textbf{0.321} &        ---        &       ---         &        ---        & \textbf{0.173} &           ---     &      ---          &       ---         \\
                      & K=10 & 0.472          & \textbf{0.547} &        ---        &        ---        & 0.340          & \textbf{0.425} &         ---       &           ---     & 0.177          & \textbf{0.239} &                ---&         ---       \\
                      & K=20 & 0.535          & 0.602          & \textbf{0.674} &      ---          & 0.364          & 0.447          & \textbf{0.537} &       ---         & 0.184          & 0.245          & \textbf{0.317} &        ---        \\
                      & K=40 & 0.614          & 0.669          & 0.726          & \textbf{0.787} & 0.389          & 0.474          & 0.562          & \textbf{0.653} & 0.197          & 0.255          & 0.324          & \textbf{0.403} \\
    \bottomrule
    \end{tabular}
\end{table*}

\subsection{Qualitative Analysis}

\begin{figure}
    \centering
    \includegraphics[width = 0.7 \linewidth]{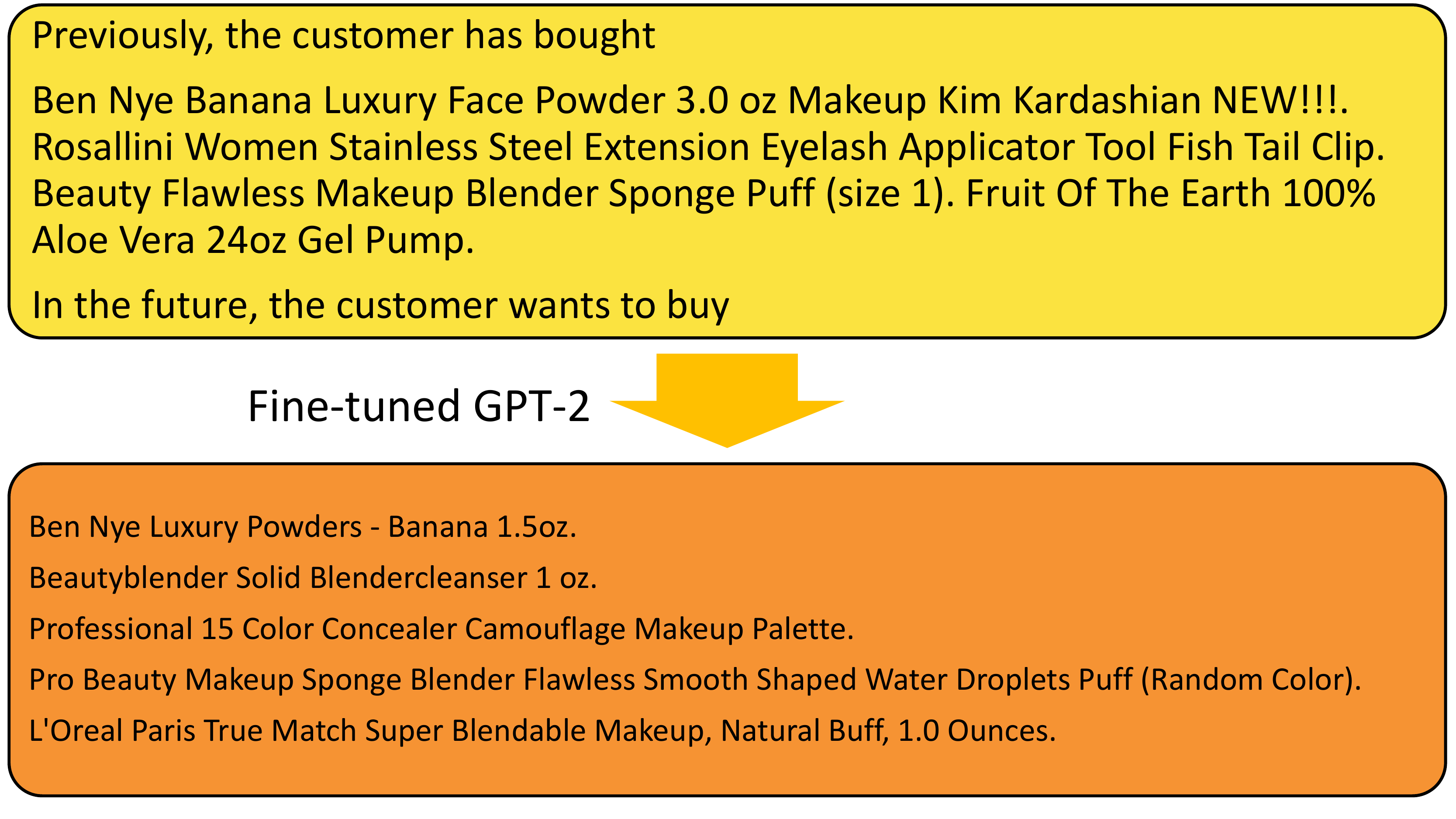}
    \includegraphics[width = 0.7 \linewidth]{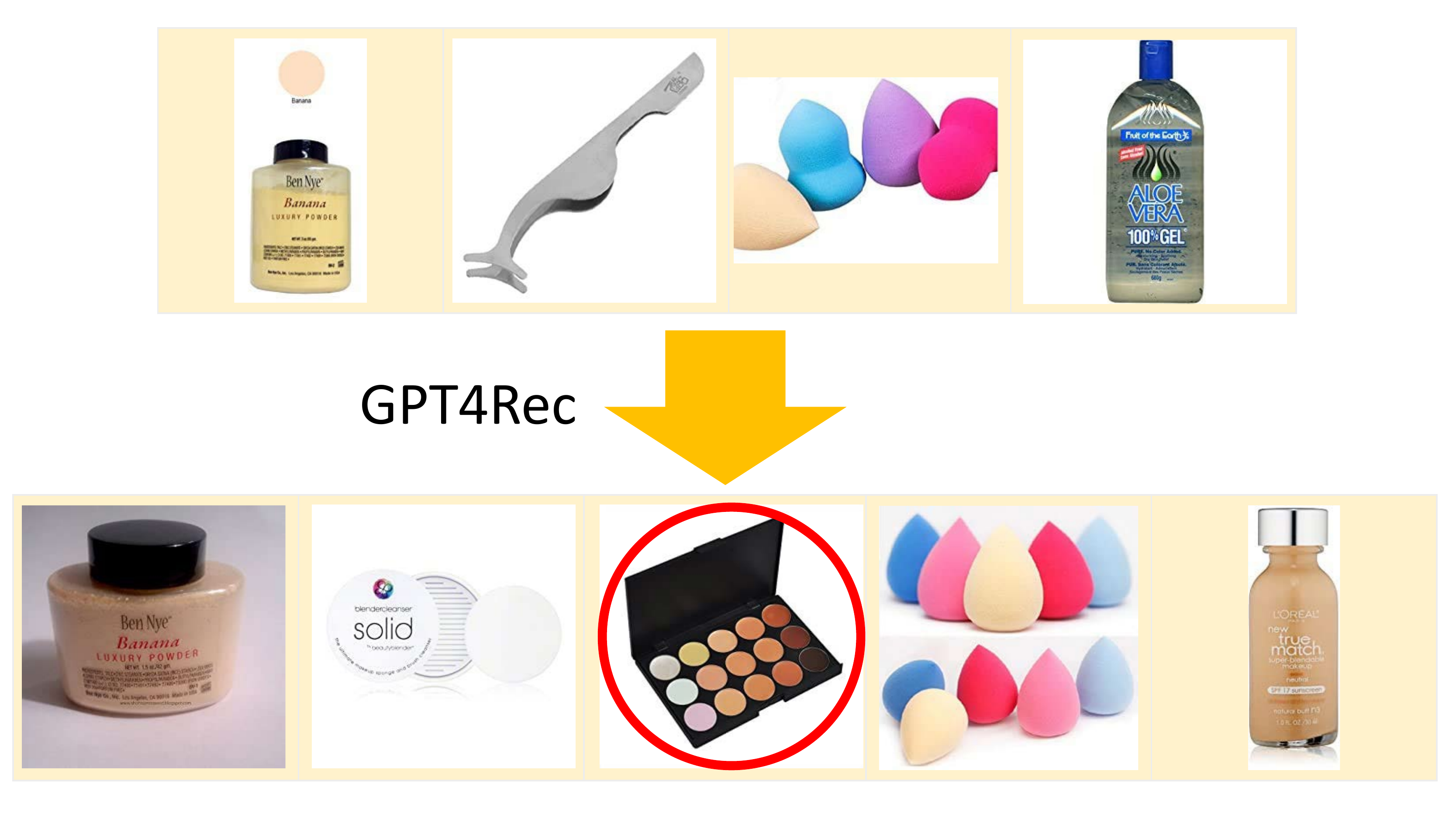}
    
    \caption{Example of a test user with diverse interests in Beauty products, with target item circled in red.}
    \label{fig:Qualitative_Example1}
\end{figure}

\begin{figure}
    \centering
    \includegraphics[width = 0.7 \linewidth]{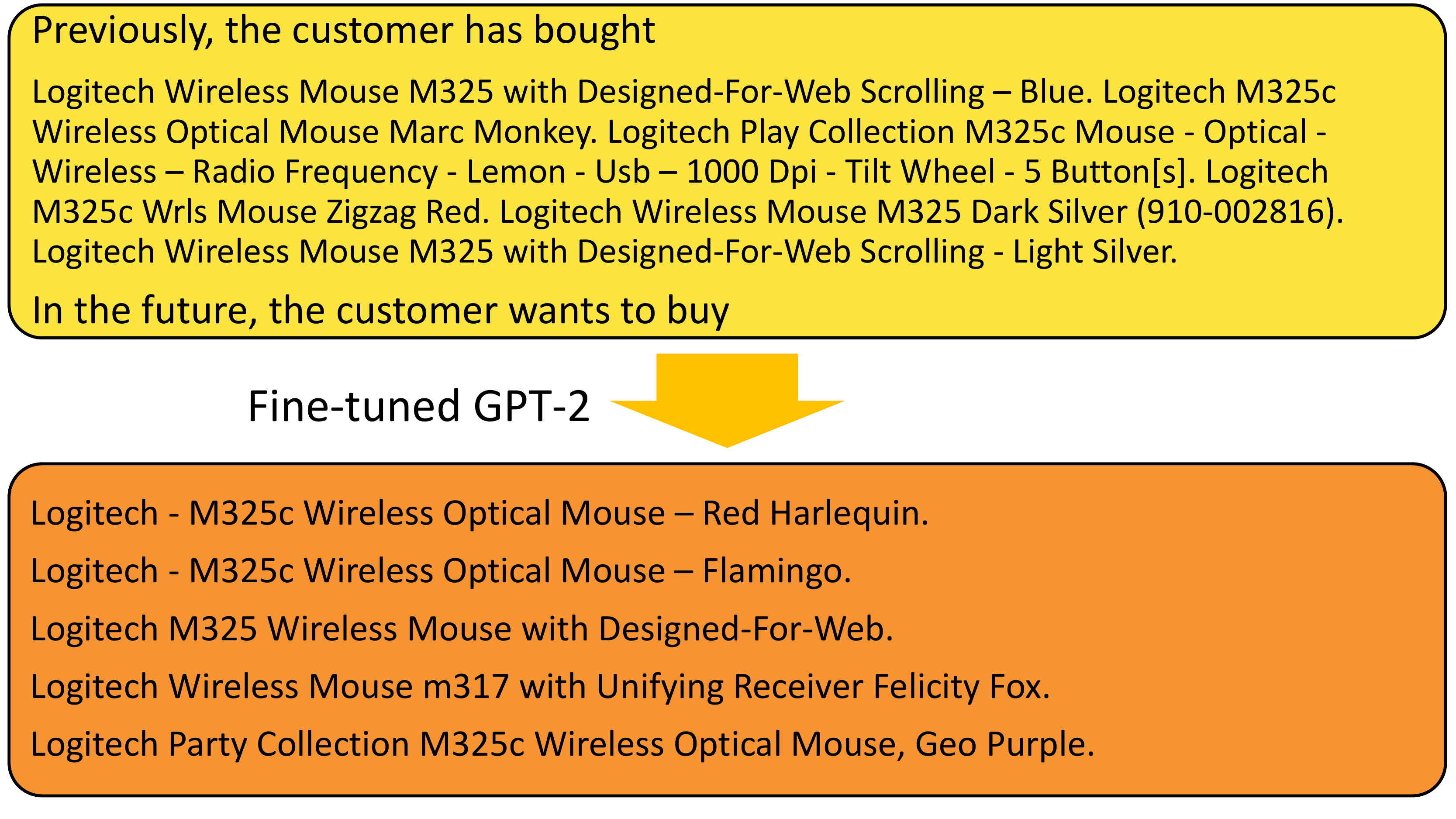}
    
    \includegraphics[width = 0.7 \linewidth]{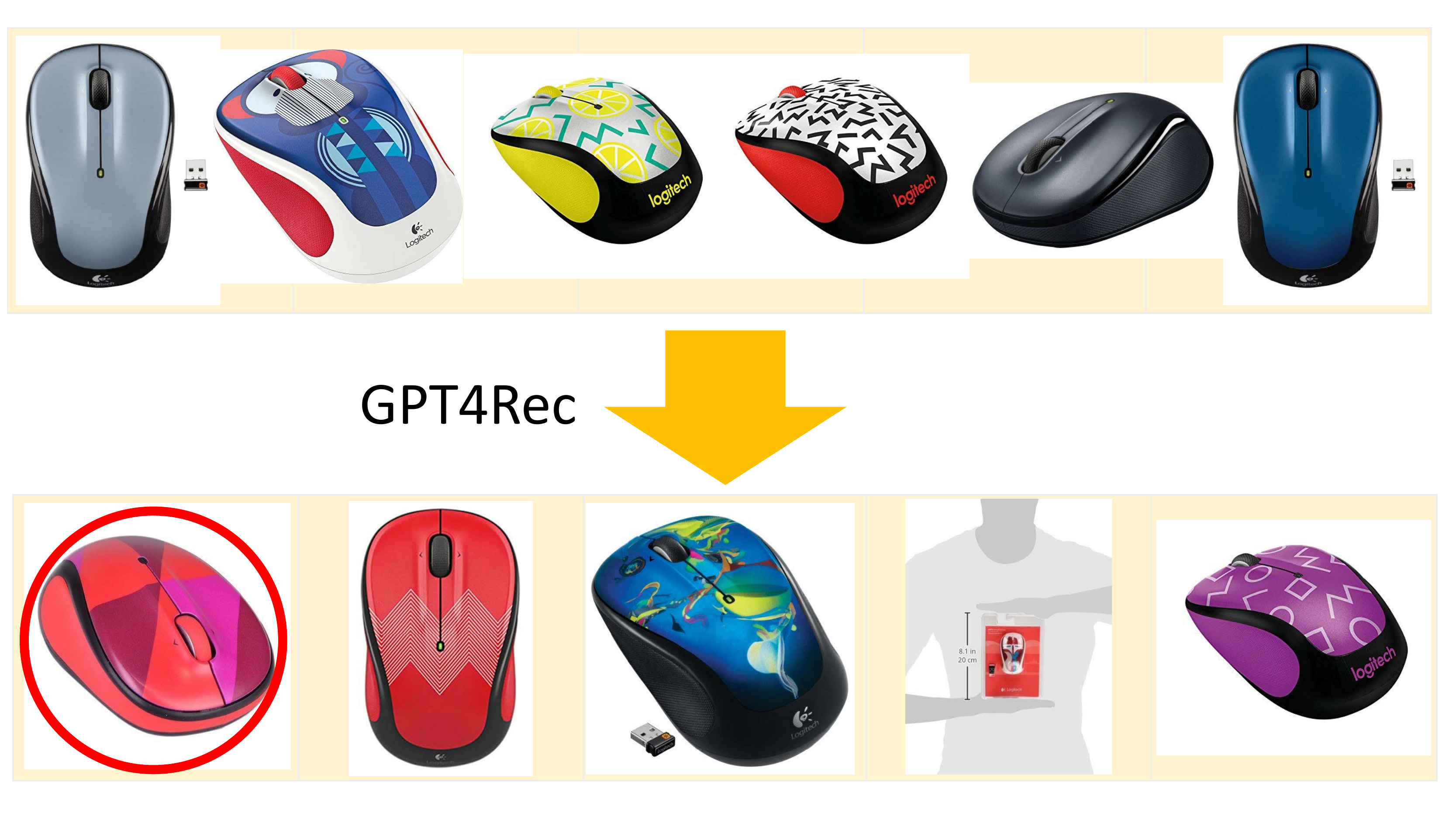}
    \caption{Example of a test user with specific interests in Electronics products, with target item circled in red.}
    \label{fig:Qualitative_Example2}
\end{figure}

In this section, we explore the effectiveness of generated queries in capturing user interests and their usefulness in interpreting user interests through case studies. We illustrate two test user sequences and model outputs in Figure \ref{fig:Qualitative_Example1} and Figure \ref{fig:Qualitative_Example2}.

In the first example, the test user has diverse interests in multiple categories and brands of Beauty products. GPT4Rec in this case is able to generate diverse queries that not only cover some products or brands in user history, but also generates relevant items that never appeared in user history such as "makeup palette". This demonstrates that GPT4Rec is able to capture the associations among items based on user-item interaction and item title semantic information. In the second example, the test user has very a specific interest in Logitech wireless mouses. GPT4Rec manages to capture this feature and all generated queries are concentrated on the same brand, same categories but with different product details.

We have demonstrated the effectiveness of GPT4Rec in generating queries that capture user interests across various aspects and granularity levels, as shown in two case studies above. These queries directly serves the purpose of interpreting user interests. Moreover, comparison of the two examples indicates the diversity level of generated queries are adaptive to the level in user sequences, which further helps understand their behavior.

%--------------------------------------------------------------------------------------------------------
%--------------------------------------------------------------------------------------------------------
\section{Conclusion}

In this work, we propose GPT4Rec, a novel generative framework that produces personalized recommendation and interpretable user interests representation simultaneously. Leveraging advanced language models and item content information, GPT4Rec achieves superior performance and naturally solves practical issues such as item cold-start problem. The proposed multi-query beam search technique generates user interests representation with different aspects and granularity, yieldings improvement in relevance and diversity of recommendation results. Our framework is flexible to incorporate more advanced generative language models or search engines, as well as better generation and retrieval strategies, which are interesting directions for future exploration.

\newpage

\bibliographystyle{ACM-Reference-Format}
\bibliography{ref}

\end{document}